A white paper for the ESO Expanding Horizons initiative

# Archaeological investigation of galaxies' evolutionary history in the cosmic middle ages

Anna R. Gallazzi[1], Stefano Zibetti[1], Mark Sargent[2], Nicolas Bouché[3], Luke Davies[4], Marcella Longhetti[5], Annagrazia Puglisi[6], Laura Scholz-Diaz[1], Fabio Ditrani[5,7], Daniele Mattolini[1,15], Sabine Thater[8], Crescenzo Tortora[9], Bodo Ziegler[8], Mirko Curti[10], Lucia Pozzetti[11], Mojtaba Raouf[12, 13], Umberto Rescigno[14]

Affiliations:
[1]INAF-Osservatorio Astrofisico di Arcetri, Largo Enrico Fermi 5, 50126, Firenze, Italy
[2]EPFL Laboratory of Astrophysics (LASTRO), Observatoire de Sauverny, CH ─ 1290, Versoix, Switzerland
[3]Univ Lyon1, Ens de Lyon, CNRS, Centre de Recherche Astrophysique de Lyon (CRAL) UMR5574, F-69230 Saint- Genis-Laval, France
[4]ICRAR, The University of Western Australia, 35 Stirling Highway, Crawley, WA 6009, Australia
[5]INAF-Osservatorio Astronomico di Brera, Via Brera 28, I-20121, Milano, Italy
[6]School of Physics and Astronomy, University of Southampton, Highfield SO17 1BJ, UK
[7]Università degli Studi di Milano-Bicocca, Piazza della Scienza, I-20125, Milano, Italy
[8]University of Vienna, Department of Astrophysics, Tuerkenschanzstrasse 17, 1180, Vienna, Austria
[9]INAF - Osservatorio Astronomico di Capodimonte, Salita Moiariello 16, I-80131 Napoli, Italy
[10]European Southern Observatory, Karl-Schwarzschild Straße 2, D-85748 Garching bei München, Germany
[11]INAF-OAS, Osservatorio di Astrofisica e Scienza dello Spazio di Bologna, via Gobetti 93/3, I-40129 Bologna, Italy
[12]Leiden Observatory, Leiden University, P.O. Box 9513, 2300 RA Leiden, Netherlands
[13]Delft University of Technology (TUDelft), Netherlands
[14]Instituto de Investigacion en Astronomıa y Ciencias Planetarias, Universidad de Atacama, Copiapo, Atacama, Chile
[15]Dipartimento di Fisica, Università di Trento, Via Sommarive 14, I-38123, Povo (TN)

**Galaxy evolution after cosmic noon: open questions in the cosmic middle ages**

**Galaxies as complex evolving systems.** The last 10 billion years of cosmic history (z<2) have seen a pervasive decline of the star formation (SF) in the Universe [1], which has led to the predominance of passive systems at the present epoch [2]. Local archaeological studies and the spectroscopic confirmation of massive evolved systems at increasingly high redshifts (z>2-3) [e.g 3, 4] consistently indicate that most of the massive galaxies (M>$10^{11}$M$_\odot$) follow a fast evolutionary path and quench early on, likely under the influence of AGN feedback. Nonetheless, they keep evolving, by changing their structure and dynamics, and by accreting satellite galaxies. Less massive galaxies (<$10^{10.5}$ M$_\odot$) maintain their SF activity on longer timescales, yet they are subject to progressive quenching in a downsizing fashion since z~2. Because of the relatively modest mass of the SMBH hosted by these galaxies, the quenching in this mass regime cannot be driven by AGN feedback but must be related to stellar/SuperNova feedback and (hydro)dynamical mechanisms, both internal and environmental, all connected in a very complex interplay. In addition to galaxy mass, *the host dark matter halo and the large-scale environment of galaxies (filaments, walls, groups and clusters) play an important role in modulating galaxy growth in tandem with cosmic structure growth*. The **cosmic "middle ages"**, from cosmic noon (z~2) to z~0.3, are thus **a key epoch to understand the drivers of galaxies' transformation when quenching is dominant and cosmic structures become mature**, and address fundamental questions such as: *what are the causes of the suppression in star formation activity? What regulates star formation and metal production efficiency over cosmic time? How do galaxies evolve after quenching their star formation? What is the relative role of global galaxy properties (such as mass) and of their large-scale environment (i.e. the cosmic web and its substructures)?*

**Stellar populations as archaeological probes of the baryon cycle.** Crucial information to answer these questions comes from a detailed characterization, across mass scales and redshift, of the physical properties of stars and gas in galaxies, which are the result of star formation and metal recycling, modulated by gas inflows and outflows and matter accretion. While the chemical composition of the ionized ISM represents the final stage of metal enrichment in star-forming galaxies and it is sensitive to short-timescale variations, *the physical properties of galaxies stellar populations (such as total stellar metallicity, element abundance ratios, IMF) are fossil records* of the integrated star-formation and assembly histories, irrespective of the ongoing SF. *The chemistry of gas and stars combined with the current SF activity and the age distribution of stellar populations can thus map different timescales of a galaxy's evolution.* All this information is **only** accessible from *deep (high S/N continuum) rest-frame optical spectroscopy, sampling with high accuracy key stellar absorption features in addition to emission line ISM diagnostics*. While SKA will survey the neutral gas content of galaxies down to a few $10^9$M$_\odot$ up to z~1, the analysis of stellar populations and of the ionized ISM will provide an essential piece of information to close the loop on our understanding of the baryon cycle.

**In the local Universe**, the unrivalled statistics and depth of the SDSS survey have established several scaling relations connecting stellar and ISM physical properties with galaxy mass down to $10^9$ M$_\odot$ [e.g. 5, 6, 7, 8, 9, 10]. The observed transition mass from young, metal-poor systems to old, metal-rich, α-enhanced populations is associated to the mass scale that regulates gas accretion mode and quenching mechanism [11,12]. Quenching mechanisms and their timescales can be further constrained by examining, at fixed stellar mass, how gaseous and stellar metallicities depend on SFR [e.g. 9, 13], and on assembly history tracers, such as size, structure and stellar dynamics [e.g. 14, 15]. Moreover, stellar populations gradients in galaxies reveal an interplay between the local and global potential well and the imprint of galaxies' accretion history [16, 17, 18].

In addition to galaxy mass, the physical properties of stellar populations in galaxies, both integrated and spatially-resolved, show subtle but significant dependences on *several environmental metrics*: galaxy hierarchy [19, 20], stellar-to-halo mass ratio [21, 22], distance from cosmic web structures [23]. Combining these different metrics and the fossil records of galaxies' SFHs from spectroscopy can help disentangle satellite-specific environmental processes from initial conditions set by the matter density field [20, 24].

**Beyond the local Universe** we still lack a solid quantification of how and when these scaling relations were established, and how their dependencies on SFR, structure and environment evolved. This is not achievable from the fossil record of present-day galaxies alone: their old stellar ages limit our ability to resolve the first few

Gyr of their evolution [25], and scaling relations are not evolutionary tracks and must be characterised at different epochs.

Constraining past SFH, stellar velocity dispersion (dark matter mass tracer), stellar metallicity and element abundance ratios (chemical enrichment tracers), and higher order kinematic moments (assembly tracers), requires *high S/N (S/N>20/Å r.f. on the continuum) on individual galaxies,* making this observationally demanding. LEGA-C [26] (~3000 K-selected galaxies at 0.6<z<1 integrated 20 hr with VIMOS/VLT) has demonstrated the potential of a comprehensive archaeological analysis, by showing e.g. that chemo-archaeological downsizing trends already existed at those epochs and evolved over the past 6-8 Gyr [e.g., 27, 28, 29]. At the same time, the high-completeness DEVILS survey reveals that photometry-based galaxy properties depend on hierarchy and halo mass at least to z~0.5 [30]. However, the small LEGA-C sample size limits robustly dissecting the scaling relations by parameters such as SFR [29] and environment [31,32]; the high stellar mass limit (>$10^{10.5}$ $M_\odot$) excludes key galaxy populations and prevents a complete census of the progenitors of typical MW-like present-day galaxies. Moreover, the LEGA-C slit spectroscopy provides limited spatially-resolved physical and kinematic information. *Questions remain on how much the inferred evolution is consistent with the decline in cosmic SFR density and the number density of galaxy populations; how much additional star formation and metal enrichment is required with respect to mass assembly through mergers or accretion*; *which environmental scales are most relevant at these redshifts.*

> To substantially advance our understanding of what regulates star formation and post-quenching galaxy evolution we must:
> - Go beyond the fundamental physical connections revealed by mean scaling relations, and **resolve their intrinsic scatter and its relation to other properties, which reflect the vast complexity of physical mechanisms**.
> - Make a **statistical yet physical connection between progenitors and descendants***,* only possible by characterizing detailed physical properties of galaxy populations at different redshifts.
>
> Therefore, *tracing the* **volume-weighted distribution of galaxies in physical parameter space across a wide range in mass, redshift and environment** *is the way forward to disentangle the interplay between assembly processes, regulated by the matter distribution in the Universe, and baryonic process, regulated by star formation, feedback and the matter cycle within galaxies.*

### Scientific requirements

A volume-complete inventory of stellar and ISM content in galaxies as a function of mass, SFR, structure and environment over the last 8-10 Gyr (z<1-2) requires *A DEEP AND EXTENSIVE SPECTROSCOPIC SURVEY* featuring:

- **Volume-representative samples** to disentangle individual galaxy evolution from population evolution (aka progenitor bias) and test cosmological models.
- **Large statistics (~$10^6$ galaxies)** to: i) sample transient/stochastic processes (mergers, inflows, outflows) including rare systems; ii) characterize the scatter in scaling relations and dissect inter-dependencies on SFR, structure, AGN, and environment across wide mass and environment ranges.
- **Accurate estimates** of stellar populations (SFH, metallicities, abundances, IMF), stellar kinematics, dynamical masses, ISM properties, and AGN diagnostics, both integrated and spatially resolved.
- **Completeness to $10^9$ $M_\odot$**, more than an order of magnitude below existing surveys, where scatter is larger, evolution stronger, and to map progenitors of typical present-day galaxies.
- **Uniform coverage at 0.3≲z≲1.5**, the cosmic middle ages bridging present-day and high-z galaxies, when intermediate-mass star formation is suppressed and environmental effects emerge.
- **Comprehensive environmental characterization** from small groups to dense clusters with sub-Mpc resolution of cosmic web structures.

*Such a survey would provide long-lasting transformative legacy value for galaxy evolution studies.*

### Technological requirements

**Galaxy physical properties:** Large statistics and high continuum S/N (S/N>20/Å rest-frame) to $10^9$ $M_\odot$ require efficient wide-area high-multiplex MOS. Extended coverage from 3700Å to J(H)-band is needed for ISM and stellar diagnostics up to z~1(z~2). Medium resolution (R~3000-5000) is essential to measure kinematics at low masses, decompose stellar continuum from ISM/AGN emission, measure absorption features, and detect outflows. A large-FoV monolithic IFU with sub-arcsec resolution would enable spatially resolved spectra for statistically significant samples, deriving maps of gas/stellar kinematics (dynamical masses) and physical properties, distinguishing outflows from merger-driven disturbances [33], and providing aperture-unbiased calibration on larger MOS samples.

**Halo masses and cosmic web:** Scales from 100 kpc to several Mpc require high spatial resolution and large-area coverage, best achieved combining wide-area high-multiplex MOS with large-FoV monolithic IFU.

These requirements demand a ***dedicated large-collecting-area spectroscopic facility with wide-area high-multiplex MOS and large-FoV monolithic IFU, equipped with medium-resolution optical-NIR spectrographs***, ideally suited for comprehensive galaxy characterization to z<1-2 in synergy with high-completeness environmental characterization.

### Limitations in the 2020-2030s landscape

The requirement of large statistics ($10^6$ galaxies) and high spectral quality (S/N>20 in the continuum) down to $\sim 10^9$ $M_\odot$ demands a large investment of telescope time, which will not be possible with current or other planned facilities. Only spectrographs with the required medium resolution (R~3000-5000) are considered here.

In the early 2030s, **WEAVE-StePS** (7hr, 25k galaxies) and **4MOST-StePS** (30hr, 3.5k galaxies), with environmental characterization from **4MOST-WAVES**, will map stellar kinematics, ages and metallicities for massive galaxies at z<0.8, but sacrifice either spectral quality or sample size, limited to $M>10^{11}M_\odot$ (I>20.5). Reaching simultaneously the required sample size, mass limit and depth is prohibitively expensive for 4m-class telescopes.

**MOONS/VLT** (1000-fiber, 500 arcmin² FoV) will explore cosmic noon with inhomogeneous samples. MOONRISE plans 8hr observations for ~60k passive galaxies (0.9<z<2.6). **PFS/Subaru** (1.25 deg², ~2400 fibers) plans 2hr integrations for ~230k galaxies at 0.7<z<1.7 ($M>10^{10.5}M_\odot$), with 14k observed for 12hr. While enabling statistical studies, their FoV and multiplexing remain inefficient for the required sample size, quality, mass limit, and volume-representativeness over a longer redshift baseline.

E-ELT will reach fainter targets, but **MOSAIC/ELT**'s limited multiplexing (~150) and small FoV (40 arcmin²) require prohibitive survey time for $>10^4$ samples. A large-survey-volume spectroscopic facility would complement ELT/ALMA follow-up where spatial resolution is critical, and would be in synergy with wide-field high-resolution optical/NIR imaging from Euclid and LSST and with HI surveys with SKA.

Moreover, all these facilities lack a large-FoV IFU component for an efficient and unbiased acquisition of large ($>10^4$) samples with spatially resolved information.